\documentstyle[psfig,graphicx]{mn}
\begin{document}
\newcommand{\lsim}{\mathrel{\rlap{\raise -.3ex\hbox{${\scriptstyle\sim}$}}%
                   \raise .6ex\hbox{${\scriptstyle <}$}}}%
\newcommand{\gsim}{\mathrel{\rlap{\raise -.3ex\hbox{${\scriptstyle\sim}$}}%
                   \raise .6ex\hbox{${\scriptstyle >}$}}}%
\def\simlt{\mathrel{\rlap{\lower 3pt\hbox{$\sim$}}
        \raise 2.0pt\hbox{$<$}}}
\def\simgt{\mathrel{\rlap{\lower 3pt\hbox{$\sim$}}
        \raise 2.0pt\hbox{$>$}}}

\title[Effect of clustering on extragalactic source counts]
{Effect of clustering on extragalactic source counts with
low-resolution instruments}

\author[M. Negrello et al.] {\parbox[t]{\textwidth}
        {M.~Negrello$^{1}$, J.~Gonz\'alez-Nuevo$^{2}$, M.~Magliocchetti$^{1}$,
        L.~Moscardini$^{3}$, G.~De Zotti$^{4,1}$, L.~Toffolatti$^{2}$, L.~Danese$^{1}$}
\vspace*{6pt} \\
$~$ \\
$^1$SISSA, via Beirut 4, I-34014 Trieste, Italy \\
$^2$Departamento de F{\'\i}sica, Universidad de Oviedo, c. Calvo
Sotelo s/n, 33007 Oviedo, Spain \\
$^3$Dipartimento di Astronomia,
Universit\`a di Bologna, via Ranzani 1, I-40127 Bologna, Italy \\
$^4$INAF -- Osservatorio Astronomico di Padova, vicolo
dell'Osservatorio 5, I-35122 Padova, Italy \\ }

\maketitle
\begin{abstract}
In the presence of strong clustering, low-resolution surveys
measure the summed contributions of groups of sources within the
beam. The counts of bright intensity peaks are therefore shifted
to higher flux levels compared to the counts of individual sources
detected with high-resolution instruments. If the beam-width
corresponds a sizable fraction of the clustering size, as in the
case of Planck/HFI, one actually detects the fluxes of clumps of
sources. We argue that the distribution of clump luminosities can
be modelled in terms of the two- and three-point correlation
functions, and apply our formalism to the Planck/HFI $850\,\mu$m
surveys. The effect on counts is found to be large and sensitive
also to the evolution of the three-point correlation function; in
the extreme case that the latter function is redshift-independent,
the source confusion due to clustering keeps being important above
the canonical $5\sigma$ detection limit. Detailed simulations
confirm the reliability of our approach. As the ratio of the
beam-width to the clustering angular size decreases, the observed
fluxes approach those of the brightest sources in the beam and the
clump formalism no longer applies. However, simulations show that
also in the case of the Herschel/SPIRE $500\,\mu$m survey the
enhancement of the bright source counts due to clustering is
important.

\end{abstract}

\section{Introduction}
\label{sec:intro}

It has long been known (Eddington 1913) that the counts (or flux
estimates) of low signal-to-noise sources are biased high.
Contributions to the noise arise both from the instrument and from
source confusion. The latter, which may dominate already at
relatively bright fluxes in the case of low-resolution surveys,
comprises the effect of Poisson fluctuations and of source
clustering. For angular scales where the correlation of source
positions is significant, the ratio of clustering to Poisson
fluctuations increases with angular scale (De Zotti et al. 1996),
so that the confusion limit can be set by the effect of
clustering. This is likely the case for some far-IR and for sub-mm
surveys from space, due to the relatively small primary apertures
and to the presence of strongly clustered populations (Scott \&
White 1999; Haiman \& Knox 1999; Magliocchetti et al. 2001;
Perrotta et al. 2003; Negrello et al. 2004).

While the effect of Poisson confusion on source counts has been
extensively investigated both analytically (Scheuer 1957; Murdoch
et al. 1973; Condon 1974; Hogg \& Turner 1998) and through
numerical simulations (Eales et al. 2000; Hogg 2001; Blain 2001),
the effect of clustering received far less attention. The
numerical simulations by Hughes \& Gazta\~{n}aga (2000) focused on
the sampling variance due to clustering. Algorithms successfully
simulating the 2D distribution of clustered sources over sky
patches as well as over the full sky have been presented by
Arg\"ueso et al. (2003) and Gonz\'alez-Nuevo et al. (2004), who
also discussed several applications. Some efforts have been made
to quantify theoretically the effect of clustering on the
confusion noise but not on the source counts (Toffolatti et al.
1998; Negrello et al. 2004; Takeuchi \& Ishii 2004).

In this paper we will use the counts of neighbours formalism
(Peebles 1980, hereafter P80) and numerical simulations to address
the effects of clustering on source fluxes as measured by
low-angular resolution surveys and, consequently, on source
counts. The formalism is described in Section~\ref{sec:formalism},
the numerical simulations in Section~\ref{sec:simulations}, while
in Section~\ref{sec:applications} we present applications to the
850$\,\mu$m survey with Planck/HFI and to the 500$\,\mu$m
Herschel/SPIRE surveys. In Section~\ref{sec:conclusions} we
summarize our main conclusions.

We adopt a flat cold dark matter (CDM) cosmology with
$\Omega_{0,\Lambda} = 0.7$ and $h=H_{0}/100\,
\hbox{km}\,\hbox{s}^{-1}\,\hbox{Mpc}^{-1}=0.7$, consistent with
the first-year WMAP results (Spergel et al. 2003).

\section{Formalism}
\label{sec:formalism}

While in the case of a Poisson distribution a source is observed on
top of a background of unresolved sources that may be either above or
below the all-sky average, in the case of clustering sources are
preferentially located in over-dense environments and one therefore
measures the sum of all physically related sources in the resolution
element of the survey (on top of Poisson fluctuations due to unrelated
sources seen in projection).

If the beam-width corresponds to a substantial fraction of the
clustering size, the observed flux is, generally, well above that
of any individual source. Thus, to estimate the counts we cannot
refer to the source luminosity function, but must define the
distribution of luminosities, $L_{cl}(z)$, of source ``clumps'',
as a function of redshift $z$. It has long been known (Kofman et
al. 1994; Taylor \& Watts 2000; Kayo et al. 2001) that a
log-normal function is remarkably successful in reproducing the
statistics of the matter-density distribution found in a number of
N-body simulations performed within the CDM framework, not only in
weakly non-linear regimes, but also in more strongly non-linear
regimes, up to density contrasts $\delta \approx 100$.
Furthermore, it displays the correct asymptotic behaviour at very
early times, when the density field evolves linearly and its
distribution is still very close to the initial Gaussian one
(Coles \& Jones 1991).

If light is a (biased) tracer of mass, fluctuations in the luminosity
density should obey the same statistics of the matter-density field;
we therefore adopt a log-normal shape for the distribution of $L_{cl}$.
Such a function is completely specified by its first (mean) and
second (variance) moments, that can be evaluated by using the counts of
neighbours formalism. The mean number of objects inside a volume $V$
centered on a given source, $\langle N
\rangle_{p}$, is [see Eq. (36.23) of P80]:
\begin{eqnarray}
\langle N \rangle_{p}=n\int_{V}[1+\xi(r)]dV\ ,
\label{eq:mean}
\end{eqnarray}
where $n$ is the mean volume density of the sources and $\xi$ is their
two-point spatial correlation function. The excess of objects (with
respect to a random distribution) around the central one is then given
by the second term on the right-hand side of this equation. The
variance around the mean value $\langle N
\rangle_{p}$ can instead be written as [Eq. (36.26) of P80]:
\begin{eqnarray}
\langle(N-\langle N \rangle_{p})^{2}\rangle_{p}= \langle N \rangle_{p}~ + ~~~~~~~~~~~~~~~~~~~~~~~~~~~~~~~~~~~ \nonumber \\
~~~~n^2\int_{V}\int_{V}[\zeta(r_{1},r_{2})+\xi(r_{12})-\xi(r_{1})\xi(r_{2})]dV_{1}dV_{2}\
. \label{eq:var}
\end{eqnarray}
If the first term on the right-hand side dominates, the variance
is approximately equal to the mean, as in the case of a Poisson
distribution.

The second term on the right-hand side of Eq.~(\ref{eq:var}) is
related to the skewness of the source distribution and exhibits a
dependence also on the reduced part of the three-point angular
correlation function, $\zeta$, for which we adopt the standard
hierarchical formula:
\begin{equation}
\zeta(r_{1},r_{2})=Q[\xi(r_{1})\xi(r_{2})+\xi(r_{1})\xi(r_{12})+
\xi(r_{12})\xi(r_{2})].
\label{eq:3ptCF}
\end{equation}
In the local Universe, observational estimates of the amplitude
$Q$ indicate nearly constant values, in the range $Q\simeq
0.6$--1.3, on scales smaller than $\sim$ 10 Mpc/h (Peebles \&
Groth 1975; Fry \& Seldner 1982; Jing \& Boerner 1998, 2004). On
larger scales, i.e. in the weakly non-linear regime where $\xi
\lsim 1$, the non-linear perturbation theory - corroborated by
N-body simulations - instead predicts $Q$ to show a dependence on
the scale (Fry 1984; Fry et al. 1993; Jing \& Boerner 1997;
Gazta\~naga \& Bernardeau 1998; Scoccimarro et al. 1998).

From Eqs.~(\ref{eq:mean}) and (\ref{eq:var}) we can derive, at a
given redshift $z$, the mean luminosity of the clump,
$\overline{L}_{cl}$, and its variance, $\sigma^{2}_{L_{cl}}$:
\begin{eqnarray}
\overline{L}_{cl}(z)&=&\overline{L}(z)  \nonumber \\
&+& \int_{\cal{L}}dL^{\prime}K(z)L^{\prime}
\Phi(L^{\prime},z)\cdot\int_{V}[1+\xi(r,z)]dV\
\label{eq:mean_Lcl}
\end{eqnarray}
and
\begin{eqnarray}
& & \!\!\!\!\!\!\!\!\!\!
\sigma^{2}_{L_{cl}}(z)=\sigma^{2}_{L}+\int_{\cal{L}}dL^{\prime}K(z)L^{\prime
2}
\Phi(L^{\prime},z)\int_{V}[1+\xi(r,z)]dV  \nonumber \\
& &
~+~\left[\int_{\cal{L}}dL^{\prime}K(z)L^{\prime}\Phi(L^{\prime},z)\right]^{2}
\cdot \nonumber \\
& & \!\!\!\!\!\!\!\!\!\!\!\!\int_{V}\int_{V}[\zeta(r_{1},r_{2},z)+
\xi(r_{12},z)-\xi(r_{1},z)\xi(r_{2},z)]dV_{1}dV_{2}.
\label{eq:var_Lcl}
\end{eqnarray}
In Eq.~(\ref{eq:mean_Lcl}) $\overline{L}(z)$ represents the mean
luminosity of the sources located at redshift $z$
\begin{equation}
\overline{L}(z)=\frac{\int_{\cal{L}}dL^{\prime}K(z)L^{\prime}
\Phi(L^{\prime},z)}{\int_{\cal{L}}dL^{\prime}K(z)\Phi(L^{\prime},z)}
\label{eq:mean_Lsources}
\end{equation}
and accounts for the fact that we are considering the resolution
elements containing at least one source. The second term in the
right-hand side of Eq.~(\ref{eq:mean_Lcl}) adds the mean
contribution of neighbours, and approaches zero as the angular
resolution increases (and correspondingly the volume $V$ decreases
as the area of the resolution element).
In both Eq.~(\ref{eq:mean_Lcl}) and Eq.~(\ref{eq:var_Lcl})
$K(z)=(1+z)L[\nu(1+z)]/L(\nu)$ is the K-correction for
monochromatic observations at the frequency $\nu$ and $\Phi$ is
the luminosity function (LF) of the sources. The range of
integration in luminosity is ${\cal L}=$[$L_{\rm min},L_{\rm
max}$], where $L_{\rm min}$ and $L_{\rm max}$ are, respectively,
the minimum and the maximum intrinsic luminosity of the sources.
The integrals over volume are carried out up to $r_{\rm max} =
3r_0$, $r_0$ being defined by $\xi(r_0)=1$.

The terms on the right-hand side of Eq.~(\ref{eq:var_Lcl}) account
for fluctuations around $L_{m}(z)$ (first term), of neighbour
luminosities (second term) and of neighbour numbers (third term).
The third term has a much steeper dependence on the angular
resolution than the second one and becomes quickly negligible as
the area of the resolution element decreases. However, as we will
show in Section~\ref{sec:applications}, it is important at the
angular resolution of Planck/HFI, implying that the distribution
of fluxes yielded by such instrument carries information on the
evolution of the three-point correlation function.

Suppose that the survey is carried out with an instrument having a
Gaussian angular response function, $f(\theta)$:
\begin{equation}
f(\theta)=e^{-(\theta/\Theta)^{2}/2}, \label{eq:response}
\end{equation}
where $\Theta$ relates to the FWHM through:
\begin{equation}
\Theta=\frac{FWHM}{2\sqrt{2\ln{2}}}\ . \label{eq:FWHM}
\end{equation}
If we adopt the usual power-law model $\xi(r)=(r_0/r)^{1.8}$ cut
at some $r_{\rm max}$, the integral of the two-point correlation
function in Eqs.~(\ref{eq:mean_Lcl}) and (\ref{eq:var_Lcl}) gives,
for a source at redshift $z$:
\begin{equation}
J_2=\int_{V}\xi(r)f(\theta)dV\simeq 25.9~
r_0^{1.8}\left[D_A(z)\Theta\right]^{1.2}\ , \label{eq:J2}
\end{equation}
where $D_A(z)$ is the angular diameter distance and we have assumed
$D_A(z)\Theta<r_{\rm max}$. This equation shows that, in the case of
strongly evolving, highly clustered sources, observed with poor
angular resolution, the mean number of physically correlated
neighbours of a galaxy falling within the beam, $nJ_2$, and therefore
their contribution to the observed flux, can be quite significant.

The probability that the sum of luminosities of sources in a clump
amounts to $L_{cl}$ is then given by:
\begin{eqnarray}
p(L_{cl},z)=\frac{
\exp\left[-\frac{1}{2}[\ln(L_{cl})-\mu_{g}(z)]^{2}/
\sigma_{g}^{2}(z)\right]}{\sqrt{2\pi\sigma_{g}^{2}(z)}L_{cl}}\ ,
\label{eq:lognormal}
\end{eqnarray}
where
\begin{eqnarray}
\mu_{g}(z) & = & \ln\left[\frac{\overline{L}^{2}_{cl}(z)}
{\sqrt{\sigma^{2}_{L_{cl}}(z)+\overline{L}^{2}_{cl}(z)}}\right], \\
\sigma^{2}_{g}(z) & = & \ln\left[\frac{\sigma^{2}_{L_{cl}}(z)}
{\overline{L}^{2}_{cl}(z)}+1\right].
\end{eqnarray}
An estimate of the ``clump'' luminosity function, $\Psi_{\rm
clump}(L_{cl},z)$, is then provided by Eq.~(\ref{eq:lognormal})
apart from the normalization factor that we obtain from the
conservation of the luminosity density:
\begin{equation}
\int\Psi_{\rm clump}(L_{cl},z)L_{cl}dL_{cl} =
\int_{\cal{L}}\Phi(L,z)LdL\ .
\end{equation}
The shift to higher luminosities of $\Psi_{\rm clump}$ compared to
$\Phi$ is then compensated by a decrease of the former function
compared to the latter at low luminosities, as low luminosity
sources merge to produce a higher luminosity clump.

If the clustering terms in Eq.~(\ref{eq:mean_Lcl}) and
Eq.~(\ref{eq:var_Lcl}) are much larger than those due to
individual sources ($L_{m}$ and $\sigma^{2}_{L}$, respectively),
the survey will detect ``clumps'' rather then individual sources,
and we can use $\Psi_{\rm clump}(L_{cl},z)$ to estimate the
counts.

Both theoretical arguments and observational data indicate that
the positions of powerful far-IR galaxies detected by SCUBA
surveys (``SCUBA galaxies'') are highly correlated (see e.g. Smail
et al. 2003; Negrello et al. 2004; Blain et al. 2004) so that
their confusion fluctuations are dominated by clustering effects.
On the contrary, Poisson fluctuations dominate in the case of the
other extragalactic source populations contributing to the
sub-millimeter counts (spiral and starburst galaxies, whose
clustering is relatively weak; cf. e.g. Madgwick et al. 2003).
Therefore we will neglect the clustering of the latter populations
(which are however included in the estimates of source counts) and
will apply the above formalism to SCUBA galaxies only; for their
evolving two-point spatial correlation function, $\xi(r,z)$, we
adopt {\it model 2} of Negrello et al. (2004). According to such
model
\begin{equation}
\xi(r,z)=b^{2}(M_{\rm eff},z)\xi_{\rm DM}(r,z),
\end{equation}
where $\xi_{DM}$ is the non-linear two-point spatial correlation
function of dark matter, computed with the recipe by Peacock $\&$
Dodds (1996; see also Smith et al. 2003), adopting a CDM spectrum
for the primordial density perturbations, with an index $n=1$, a
shape parameter $\Gamma=0.2$ and a normalization $\sigma_8=0.8$
(see e.g. Lahav et al. 2002; Spergel et al. 2003); $b(M_{\rm
eff},z)$ is the redshift-dependent (linear) bias factor (Sheth
$\&$ Tormen 1999), $M_{\rm eff}$ being the effective mass of the
dark matter halos in which SCUBA sources reside. Following
Negrello et al. (2004) we set  $M_{\rm eff}=1.8\times10^{13}\,
M_{\odot}$/h, which yields values of $\xi(r,z)$ consistent with
the available observational estimates. Note that a one-to-one
correspondence between haloes and sources has been assumed.

For the amplitude, $Q$, of the three-point angular correlation
function, $\zeta$ [Eq.~(\ref{eq:3ptCF})], we consider three
models:
\begin{itemize}
\item[(i)] $Q(z)=Q(0) = 1$,
\item[(ii)] $Q(z)=Q(0)/b(M_{\rm eff},z)=1/b(M_{\rm eff},z)$,
\item[(iii)] $Q(z)=Q(0)/b^{2}(M_{\rm eff},z)=1/b^{2}(M_{\rm eff},z)$,
\end{itemize}
where we have neglected any dependence of $Q$ on scale. Both
calculations based on perturbation theory (e.g. Juszkiewicz,
Bouchet $\&$ Colombi 1993; Bernardeau 1994) and N-body simulations
(e.g. Colombi, Bouchet $\&$ Hernquist 1996; Szapudi et al. 1996)
suggest that model (i) applies to dark matter. On the other hand,
the three-point correlation of luminous objects decreases as the
bias factor $b$ increases (Bernardeau \& Schaeffer 1992, 1999;
Szapudi et al. 2001); the formula (ii), derived from perturbation
theory, is expected to hold on scales $\simgt 10$ Mpc/h (Fry $\&$
Gazta\~naga 1993) while for scales smaller than these, Szapudi et
al. (2001) quote model (iii) as a phenomenological rule derived
from N-body simulations. A more realistic model should allow for
the dependence of $Q$ on the linear scale, induced by the
increasing strength of non-linear effects with decreasing scale.
The models (i)-(iii) may thus bracket the true behaviour of
$Q(z)$, which should, however, be not far from model (iii) for the
scales of interest here.

\begin{figure}
 \vspace{7.1cm} \includegraphics{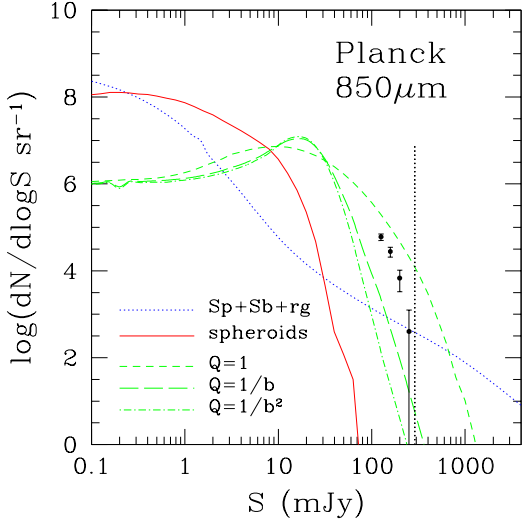} \caption{Differential source counts
 of SCUBA galaxies at 850$\,\mu$m (solid line) compared with counts
 expected in the case of observations performed with the angular
 resolution of Planck/HFI for the three models for the evolution of the
 amplitude of the three-point correlation function, $Q(z)$, (model (i):
 short dashes; model (ii): long dashes; model (iii): dot-dashed
 line). The summed contributions from (unclustered) spiral
 galaxies, starburst galaxies, and extragalactic radio sources
 is represented by the dotted line. The filled circles with error bars show
 the counts estimated from the simulations (see text). The vertical
 dotted line shows the Planck/HFI $5\sigma$ detection limit
 ($S_d=288\,$mJy) estimated by Negrello et al. (2004) allowing for
 clustering fluctuations.}  \label{fig:dNdS_Planck}
\end{figure}

\section{Numerical simulations}
\label{sec:simulations}

To test the reliability of our analytical approach, numerical
simulations of sky patches were performed using the fast algorithm
recently developed by Gonz{\' a}lez-Nuevo et al. (2004). Only SCUBA
galaxies were taken into account. Sources were first randomly
distributed over the patch area, with surface densities given, as a
function of the flux density, by the model of Granato et al. (2004);
we have considered sources down to $S_{\rm min}= 0.01\,$mJy. Then, the
projected density contrast as a function of position, $\delta({\mathbf
x})$, was derived and its Fourier transform, $\delta({\mathbf k})$,
was computed. Next, in Fourier space, we added the power spectrum
corresponding to the two-point angular correlation function,
$w(\theta)$, given by model 2 of Negrello et al. (2004), to the white
noise power spectrum corresponding to the initial spatial distribution
and obtained the transformed density field of spatially correlated
sources. Then we apply the inverse Fourier transform to get the
projected distribution of clustered sources in the real space and we
associate randomly the fluxes to the simulated sources according to
the differential counts predicted by the model of Granato et
al. (2004) (for more details, see Gonz{\' a}lez-Nuevo et al. 2004)

Simulations were carried out for surveys in:
\begin{itemize}
\item[$\triangleright$] the 850$\,\mu$m channel of the High Frequency
Instrument (HFI) of the ESA Planck satellite
(FWHM=300$^{\prime\prime}$; Lamarre et al. 2003);
 \item[$\triangleright$] the 500$\,\mu$m channel of the Spectral and
 Photometric Imaging REceiver (SPIRE) of the ESA Herschel
 satellite (FWHM=34.6$^{\prime\prime}$; Griffin et al. 2000).
\end{itemize}
In the Planck/HFI case, we have simulated sky patches of
$12^{\circ}.8\times 12^{\circ}.8\,\hbox{deg}^2$ with pixels of
$1.5\times 1.5\,\hbox{arcmin}^2$. In the Herschel/SPIRE case the
patches were of $3^{\circ}.28\times 3^{\circ}.28\,\hbox{deg}^2$
and the pixel size was $1/3$ of the FWHM.

To check the simulation procedure we have compared the confusion
noise, $\sigma$, obtained from them with the analytical results by
Negrello et al. (2004). In the case of Planck/HFI, from the
simulations we get $\sigma_{P}= 20\,$mJy and $\sigma_{C}= 50\,$mJy
for Poisson distributed and clustered sources, respectively, in
very good agreement with the analytic results, $\sigma_{P}=
20\,$mJy and $\sigma_{C}= 53\,$mJy. For the Herschel/SPIRE channel
the simulations give $\sigma_{P}= 6.4\,$mJy and $\sigma_{C}=
1.2\,$mJy, to be compared with $\sigma_{P}= 6.0\,$mJy and
$\sigma_{C}= 3.2\,$mJy obtained by Negrello et al. (2004). The
apparent discrepancy of the results for $\sigma_{C}$ actually
corresponds to the uncertainty in this quantity, very difficult to
determine from the simulations when Poisson fluctuations dominate.

Note that for the very steep counts predicted by the Granato et
al. (2004) model, that accurately matches the SCUBA and MAMBO
data, both the Poisson and the clustering fluctuations, for the
angular resolutions considered here, are independent of the flux
limit. We have checked that the results obtained from simulations
are independent of the pixel size used.

\begin{figure*}
 \includegraphics[width=0.9\columnwidth]{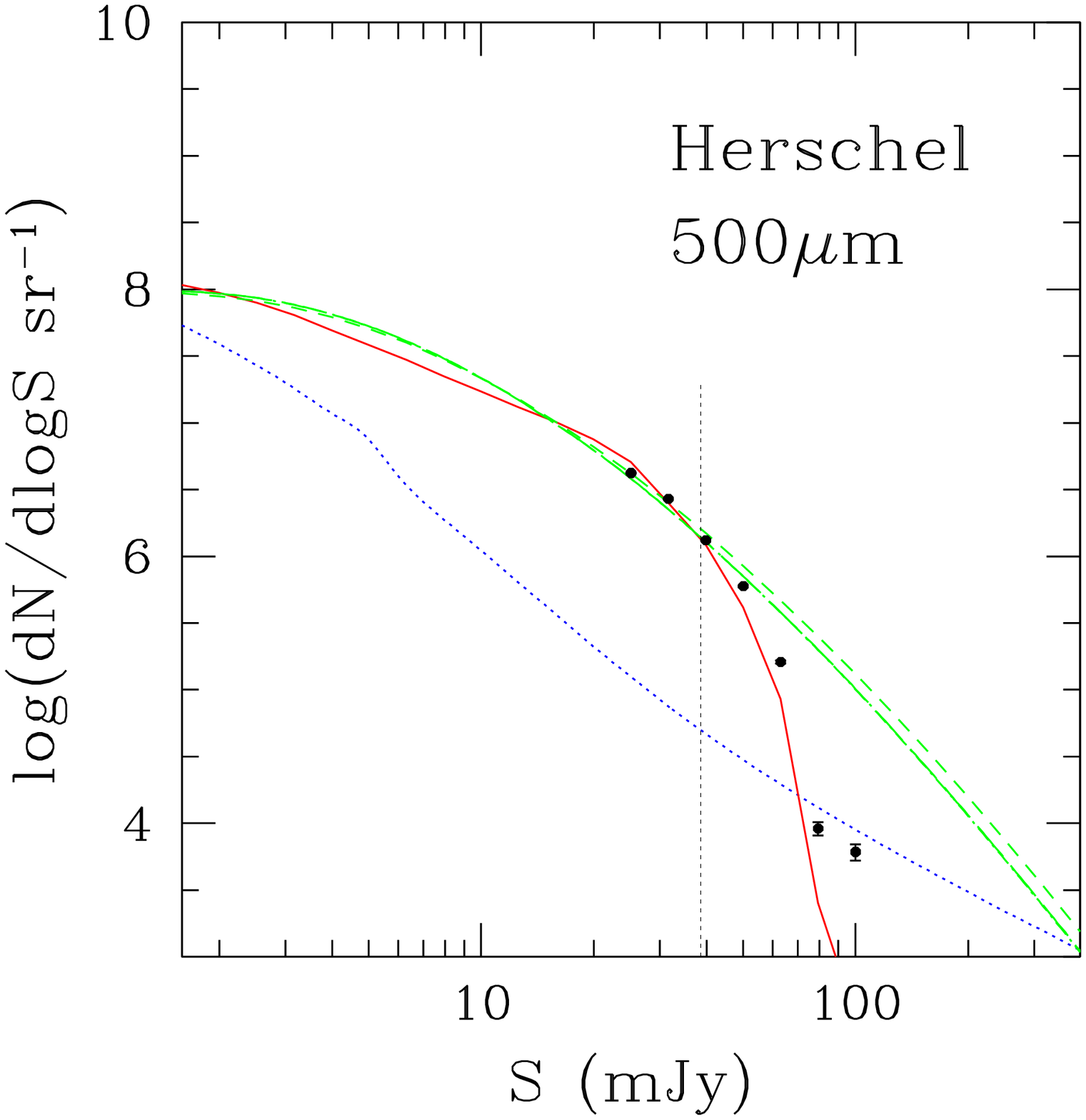}\hspace{0.3
 cm} \includegraphics[width=0.9\columnwidth]{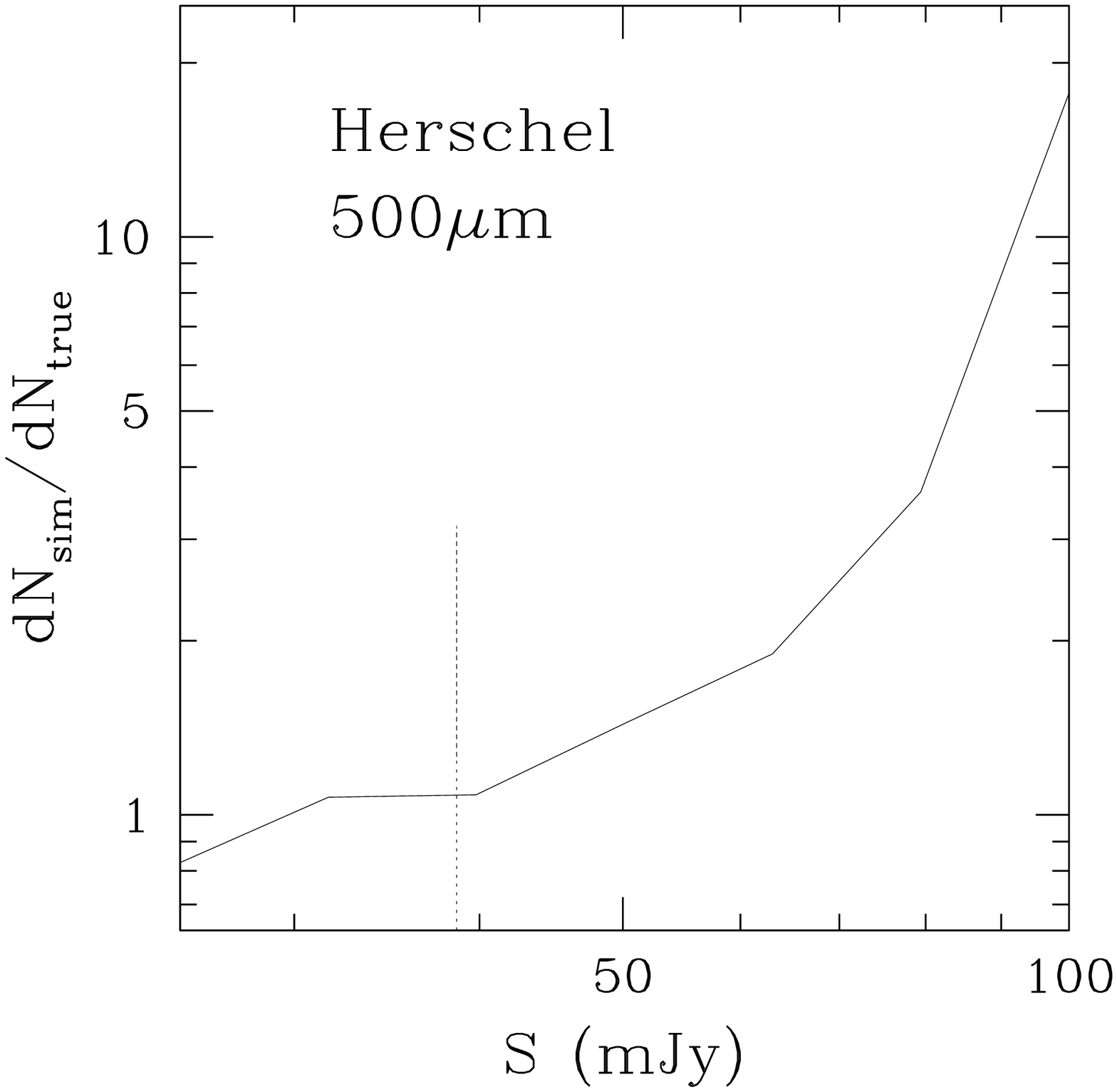}
 \caption{{\it Left-hand panel}: same as in Fig.~\protect\ref{fig:dNdS_Planck}
 but for the 500$\,\mu$m channel of Herschel/SPIRE. The results for
 models (ii) and (iii) for the evolution of $Q(z)$ overlap, while model (i)
 is only slightly higher. {\it Right-hand panel}: ratio between
 the differential counts of SCUBA galaxies estimated from the
 simulations for the Herschel/SPIRE resolution and
 those predicted by the model of Granato et
 al. (2004). The dotted vertical line, in both panels,  corresponds to the $5\sigma$
detection limit
 ($S_d=39\,$mJy) estimated by Negrello et al. (2004) accounting also
 for clustering fluctuations.  }  \label{fig:dNdS_Herschel}
\end{figure*}

\section{Results and discussion}
\label{sec:applications}
The results for the Planck/HFI 850$\,\mu$m channel are shown in
Fig.~\ref{fig:dNdS_Planck}, where the solid line gives the counts
of SCUBA galaxies predicted by the physically grounded model of
Granato et al. (2004), and the dotted line gives the summed counts
of spiral and starburst galaxies, and of extragalactic radio
sources. For the redshift-dependent luminosity functions of spiral
and starburst galaxies we have adopted the same phenomenological
models as in Negrello et al. (2004), while for radio galaxies we
have used the model by De Zotti et al. (2004). The short-dashed,
long-dashed, and dot-dashed lines show the counts of ``clumps''
expected from our analytic formalism for the three evolution
models of the amplitude $Q$ of the three-point correlation
function. As expected, at the Planck/HFI resolution the counts of
``clumps'' are sensitive to the evolution of the three-point
correlation function. Clearly the predicted counts below $\simeq
50\,$mJy, where Poisson fluctuations become important, are of no
practical use; they are shown just to illustrate how the formalism
accounts for the disappearance of lower luminosity objects which
merge into the ``clumps''.

The filled circles with error bars in Fig.~\ref{fig:dNdS_Planck}
are obtained filtering the simulated maps with a Gaussian response
function of 5$^\prime$ FWHM to mimic Planck/HFI observations. The
lower flux limit of the estimated counts is set by Poisson
fluctuations. It should be noted that the procedure used for the
simulations takes into account only the two-point angular
correlation function and does not allow us to deal with the
three-point correlation function and with its cosmological
evolution, which are included in the analytic model. In principle
it is possible to go the other way round, i.e. to evaluate from
the simulations the reduced angular bispectrum, $b_{r}$, by
applying the standard Fourier analysis (Gonz{\' a}lez-Nuevo et al.
2004; Arg{\" u}eso et al. 2003) and infer from it an estimate of
the three-point correlation function weighted over the redshift
distribution. However, the relationship of the three-point
correlation function with the angular bispectrum is through a six
dimensional integral which is really difficult to deal with in
practice (Szapudi 2004). An alternative possibility consists in
computing the number of triplets (above a fixed flux threshold)
and in using the estimator developed by Szapudi $\&$ Szalay (1998)
to derive an estimate of the three-point angular correlation
function. On the whole, these methods turn out to be impractical
to take into account also the effect of the evolving three-point
correlation function when comparing simulations with analytic
results. On the other hand, it is very reassuring that the counts
obtained from simulations are within the range spanned by analytic
models.

In the Herschel case, the beam encompasses only a small fraction
of the ``clump'' and therefore the observed flux is generally
dominated by the single brightest source in the beam. Thus, the
analytic model described in Section~\ref{sec:formalism} is no
longer applicable. As illustrated by the left-hand panel of
Fig.~\ref{fig:dNdS_Herschel}, such formalism would strongly
over-predict the observed counts at bright flux densities (and the
results are essentially independent of the three-point correlation
function). On the other hand, the simulations (filled circles with
error bars) show that neighbour sources appreciably contribute to
the observed fluxes, hence to the counts at bright flux density
levels, as more clearly illustrated by the right-hand panel of
Fig.~\ref{fig:dNdS_Herschel}. Such count estimates were obtained
filtering the simulated maps with a Gaussian response function of
FWHM=34.6$^{\prime\prime}$, appropriate for the Herschel
$500\,\mu$m channel; again, the lower flux limit of the estimated
counts is set by Poisson fluctuations.

\section{Conclusions}
\label{sec:conclusions}

Theoretical arguments and observational data converge in
indicating that the very luminous (sub)-mm sources detected by
SCUBA and MAMBO surveys are highly clustered (clustering radius
$r_0 \simeq 8\hbox{h}^{-1}\,$Mpc). On the other hand, the limited
sizes of telescopes of forthcoming space instruments operating in
the sub-mm domain, such as Planck/HFI and Herschel/SPIRE, imply
relatively poor angular resolutions ($\hbox{FWHM} \simeq 5'$ for
Planck at $850\,\mu$m, and $\simeq 34.6^{\prime\prime}$ for
Herschel at at $500\,\mu$m).

In the Planck/HFI case the summed fluxes of physically correlated
sources within the beam are generally higher than the luminosity
of the brightest source in the beam, so that the outcome of the
surveys will be counts of ``clumps'' of sources rather than of
individual sources. We have argued that the luminosity
distribution of such ``clumps'' at any redshift can be modelled
with a log-normal function, with mean determined by the average
source luminosity and by the average of summed luminosities of
neighbours, and variance made of three contributions (the
variances of source luminosities, of neighbour luminosities, and
of neighbour numbers). The latter contribution depends on the
three-point correlation function, so that the counts of ``clumps''
provide information on this elusive quantity and on its
cosmological evolution. Under the, rather extreme, assumption that
the coefficient, $Q$, of the three-point correlation function is
independent of redshift, the counts of ``clumps'' extend beyond
the formal $5\sigma$ detection limit. In the more likely cases of
$Q$ decreasing as $b^{-1}$ or $b^{-2}$, $b$ being the bias factor,
the ``clumps'' will only show up in Planck maps as $< 5\sigma$
fluctuations. Anyway, the Planck surveys will provide a rich
catalogue of candidate proto-clusters at substantial redshifts
(typically at $z\simeq 2$--3), very important to investigate the
formation of large scale structure and, particularly, to constrain
the evolution of the dark energy thought to control the dynamics
of the present day universe. Detailed numerical simulations
carried out using the fast algorithm recently developed by Gonz{\'
a}lez-Nuevo et al. (2004) are fully consistent with the analytic
results, although a full comparison would require an upgrade of
the algorithm to include the effect of the evolving three-point
correlation function.

As the ratio of the beam-width to the clustering angular size
decreases, the observed fluxes approach those of the brightest
sources in the beam and the ``clump'' formalism no longer applies.
However, simulations show that also in the case of the
Herschel/SPIRE $500\,\mu$m survey the contribution of neighbours
to the observed fluxes enhances the bright tail of the observed
counts. Due to the extreme steepness of such tail, as predicted by
the model of Granato et al. (2004), even a modest addition to the
fluxes of the brightest sources may lead to counts at flux
densities $\simeq 100\,$mJy several times higher than would be
observed with a high resolution instrument. It should be noted
that, in the case of strong clustering, the canonical $5\sigma$
detection limit (shown by the vertical dotted line in both panels
of Fig.~\ref{fig:dNdS_Herschel}) does not frees the observed
counts from the confusion bias.

\noindent
\section*{ACKNOWLEDGMENTS}
We acknowledge very useful suggestions from the referee and from the
editor, that greatly helped to overcome a serious weakness of the
approach used in a previous version of this paper. We are also
indebted to G.L. Granato and L. Silva for having provided, in a
tabular form, the redshift-dependent model luminosity functions of
SCUBA galaxies at 500 and 850 $\,\mu$m. JGN and LT acknowledge partial
financial support from the Spanish MCYT under project
ESP2002-04141-C03-01. JGN acknowledges a FPU fellowship of the Spanish
Ministry of Education (MEC). Work partially supported by ASI and
MIUR.

\end{document}